\documentclass[amsmath,amssymb,aps,prl,showpacs,10pt,twocolumn]{revtex4-1}

\usepackage{dcolumn}
\usepackage{graphicx}

\begin{document}

\title{Controllable binding of polar molecules and meta-stability of 1-D gases 
       with attractive dipole forces}

\author{Jason N. Byrd}
\email{byrd@phys.uconn.edu}
\author{John A. Montgomery, Jr.}
\author{Robin C\^{o}t\'{e}}
\affiliation{Department of Physics, University of Connecticut, Storrs, CT 06269}

\begin{abstract}
We explore one-dimensional (1-D) samples of ultracold polar
molecules with attractive dipole-dipole interactions and show 
the existence of a repulsive barrier due to a strong quadrupole interaction
between molecules. This barrier can stabilize a gas of ultracold KRb molecules 
and even lead to long-range wells supporting bound states between molecules. 
The properties of these wells can be controlled by external electric fields, 
allowing the formation of long polymer-like chains of KRb, and studies of 
quantum phase transitions by varying the effective interaction between 
molecules. We discuss the generalization of those results to other systems.
\end{abstract}

\pacs{ }

\maketitle

The recent achievements in the formation and manipulation of ultracold polar
molecules \cite{miranda2011,deiglmayr2011} have opened the gate to exciting new
studies in several fields of physical sciences. Polar molecules could find uses
in quantum information \cite{yelin2006} and precision measurements
\cite{demille2000}, while their long-range and anisotropic interactions in dense
samples could provide a fertile ground for novel quantum gases
\cite{santos2000}.  In addition, advances in controlling the alignment and
orientation of polar molecules \cite{nielsen2012,holmegaard2009} enable the
manipulation of these inter-molecular interactions, building a bridge between
atomic, molecular, and optical (AMO) physics, physical chemistry, and condensed
matter physics.  Until now, stable dipolar gases were thought to require a
repulsive dipole-dipole interaction, such as provided by parallel dipoles
perpendicular to a 2-D plane.  However, to observe interesting new correlations
and phases, such as the Luttinger liquid transition \cite{recati2003} attractive
interactions are needed.  In this work, a system with such features is proposed
and investigated, combining available techniques to produce ultracold polar
molecules with the ability to precisely control their spatial orientation.
 
\begin{figure}[b]
 \caption{ \label{fig1-letter} KRb+KRb PES for coplanar geometries: aligned
 (top), T-oriented (middle), and collinear (bottom).  The inset sketches the
 geometry: ${\bf R}$ joins the geometric center two KRb, $\theta_1$ and
 $\theta_2$ are the angles between their molecular axes and ${\bf R}$, and
 $\phi$ is the angle between the molecular planes. }
 \includegraphics[width=\columnwidth]{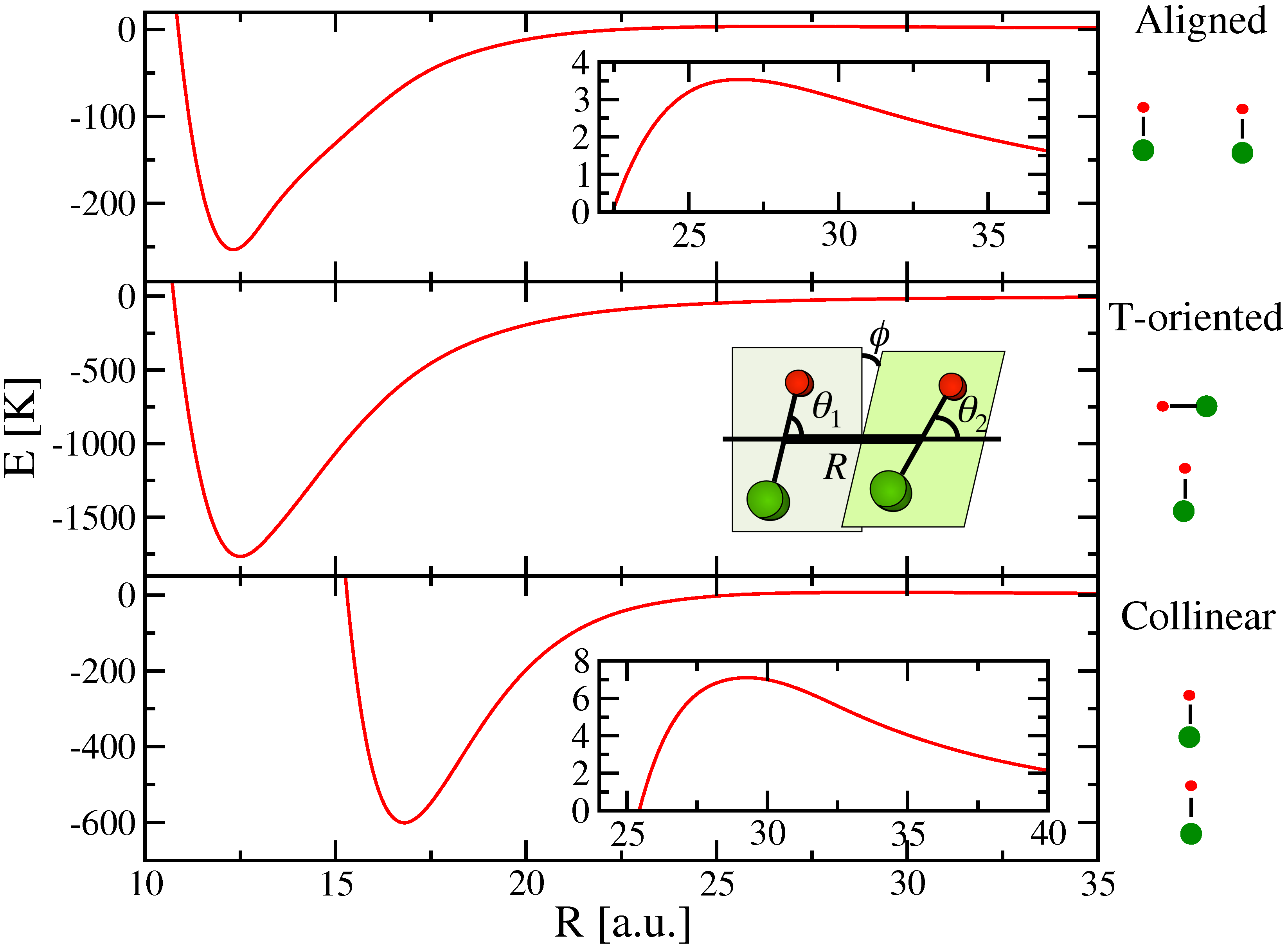} 
\end{figure}

In this work, we focus our attention on KRb, which has been trapped in
relatively large amounts \cite{miranda2011}. We first calculate the potential
energy surface (PES) $V(R,\theta_1,\theta_2,\phi)$ of two KRb molecules
approaching each other for a wide range of geometries.  We assume that both
molecules are in the ro-vibrational ground state of their electronic
X$^1\Sigma^+$ ground state, and rigid rotors, an approximation that is valid for
$R\sim 20$ a.u. (with the bond stretching by less than 0.15\%) or larger.
Fig.~\ref{fig1-letter} shows the PES for three particular geometries when both
molecular axes are in the same plane ($\phi=0$): the top panel depicts $V$ when
the molecules are aligned ($\theta_1=\theta_2=90^{\rm o}$), the middle panel for
the $T$-orientation ($\theta_1=0,\theta_2=90^{\rm o}$), and the bottom panel for
collinear molecules ($\theta_1=\theta_2=0$). Those curves illustrate the
difference between the stronger short-range region where the electronic
wavefunction becomes perturbed and the weaker long-range region where the bond
length of each KRb is not affected. The short-range region is generally deep,
with wells that depend strongly on the particular geometry, ranging from a few
100 K in Fig.~\ref{fig1-letter} for co-planar geometries, to the tetramer
K$_2$Rb$_2$ bound by $\sim 4300$ K with respect to the KRb+KRb threshold
\cite{byrd2010}.

The KRb$+$KRb PES was calculated at the
CCSD(T) level of theory using MOLPRO 2009.1 \cite{molpro09short,knowles1993},
with the K and Rb core electrons replaced by the Stuttgart relativistic ECP18SDF
\cite{fuentealba1982} and ECP36SDF \cite{fuentealba1983} pseudopotentials,
respectively.  The core-core and core-valence correlation energy was modeled
using a core polarization potential \cite{fuentealba1982}.  Supplemental basis
functions were added to existing basis sets for K \cite{magnier1996} and
Rb \cite{aymar2005}.  Uncontracting the basis sets, the exponents were 
optimized to reproduce the experimental equilibrium bond length, $R_e$, 
and dissociation energy, $D_e$ \cite{pashov2007}.  In the long-range
region where the interaction is small and the wave function
overlap between the two molecules is negligible, the interaction can be 
split into electrostatic and dispersion contributions.

Our analysis is concentrated on the coplanar geometries of Fig.~\ref{fig1-letter},
which depict a seemingly surprising result. While the top and middle 
panels depict the expected behavior of a repulsive and slightly attractive 
dipole-dipole interaction, respectively, the collinear geometry (bottom panel)
reveals a barrier. The existence of this barrier can be traced to a strong 
repulsive quadrupole interaction (see below). We also notice that it is 
higher (almost 7 K in height) than that of the aligned geometry (about 4 K). 
To better understand these {\it ab initio} 
results, we examine the KRb+KRb interaction at large intermolecular separation $R$
{\it via} the long-range expansion 
\begin{equation}
\label{eq:expansion-general}
   V(R,\theta_1,\theta_2,\phi) \stackrel{R \mbox{ \scriptsize large}}{=}
   -\sum_n \frac{W_n(\theta_1,\theta_2,\phi)}{R^n}\; . 
\end{equation} 
The functions $W_n$ may contain electrostatic ({\it e.g.} dipole ${\cal D}$, 
quadrupole ${\cal Q}$, octupole ${\cal O}$, or higher-order 
moments) and/or dispersion and induction contributions 
$C_{n,i}$ \cite{mulder1979}.  The first few terms (up to $n=6$) are
\begin{eqnarray*}
   W_3 \!& \!=\! &\! {\cal D}^2 \left( 2 c_1 c_2 - s_1 s_2 c_\phi \right) , \\
   W_4 \!& \!=\! &\!  \frac{3{\cal DQ}}{2} \left( 1 + 3 c_1 c_2 
                      - 2 s_1 s_2 c_\phi \right) \left( c_1 - c_2 \right) , \\
   W_5 \!& \!=\! &\!  {\cal D O} \biggl\{ \frac{3}{2} s_1 s_2 c_\phi
             ( 2 \!-\! 5 c^2_1 \!-\! 5 c^2_2 ) \!-\! c_1 c_2
             (6 \!-\! 5 c^2_1 \!-\! 5 c^2_2 ) \biggr\} \nonumber \\
       &  &    \hspace{-.3in} - {\cal Q}^2 \biggl\{ \frac{3}{2} (1\!-\!3 c^2_1)(1\!-\!3 c^2_2) 
             \! -\! 12 c_1 c_2  s_1 s_2 c_\phi \!+\! \frac{3}{4} s^2_1 s^2_2 c_{2\phi}
                           \biggr\} ,\\
   W_6 \!& \!=\! &\!  C_{6,0} + C_{6,1}(3 c^2_1 \!+\! 3 c^2_2 \!-\!2 ) 
                  + C_{6,2}(3 c^2_1 \!-\!1)(3 c^2_2 \!-\!1 ) \nonumber \\
       &   & + C_{6,3} c_1 c_2  s_1 s_2 c_\phi 
                + C_{6,4} s^2_1 s^2_2 c_{2\phi}
\end{eqnarray*}
where $c_i \equiv \cos\theta_i$, $s_i\equiv\sin\theta_i$, $c_{k\phi} \equiv \cos
k\phi$.  In Table ~\ref{tab1}, we list the corresponding parameters 
obtained by least squares fit of the
PES up to $n=8$.
The fitted ${\cal D}$, ${\cal Q}$, and ${\cal O}$ 
are also compared to {\it ab initio} values calculated at the all electron CCSD level 
of theory with the Roos ANO basis set \cite{roos2003}. ${\cal D}$ 
and ${\cal Q}$ agree to better than 1\%,  
attesting to the accuracy of the PES, while ${\cal O}$ is
off by one order of magnitude, reflecting the difficulty of fitting the small 
contribution of ${\cal D O}$ compared to that of ${\cal Q}^2$ in $W_5$
(${\cal D} \ll {\cal Q}$); ${\cal O}$ does not play a significant role for KRb.
Using Eq.(\ref{eq:expansion-general}), one can easily
understand the physical origin of the barriers.  For parallel
molecules, {\it i.e.} $\theta_1 = \theta_2 \equiv \theta$ and $\phi=0$, the two
leading terms in $V$ are 
\begin{equation} 
   \label{eq:V-approx}
   V(R,\theta) \simeq -\frac{W_3}{R^3} - \frac{W_5}{R^5} \; .
\end{equation}
For collinear KRb, $\theta = 0$, with $W_3 = 2 {\cal D}^2$ and $W_5 = -6 {\cal
Q}^2 + 4 {\cal D} {\cal O} \simeq - 6 {\cal Q}^2$, and because of the relatively
weak ${\cal D}$ when compared to ${\cal Q}$, the long-range attractive $R^{-3}$
dipole interaction is overcome by a shorter-range repulsive $R^{-5}$ quadrupole
interaction (the attractive contribution of ${\cal D O}$ is much weaker than
that of the repulsive ${\cal Q}^2$); at shorter range still, the attractive
$R^{-6}$ and higher contributions dominate (mostly due to the isotropic $C_{6,0}$ term) 
and bring $V$ down, hence the barrier.  
For aligned KRb, $\theta = 90^{\rm o}$, with
$W_3 = -{\cal D}^2$ and $W_5 = -(9/4) {\cal Q}^2 + 3 {\cal D} {\cal O} \simeq -
(9/4) {\cal Q}^2$, and although both leading contributions are repulsive, the
leading repulsive $W_5$ is about 3 times smaller than for the collinear case,
hence the smaller barrier shown in Fig.~\ref{fig1-letter} (the leading
attractive $C_{6,0}$ term is the same in both cases).  

\begin{table}[t]
\begin{tabular}{cc|cccccccc} \hline\hline
\multicolumn{2}{c|} {KRb fit} 
    & AB & $R_e$ & ${\cal D}$ & ${\cal Q}$ & ${\cal O}$ 
    & $W_6$ & $R_{\rm sr}$ & $R_{\cal Q}$
    \\ \hline
${\cal D}$ & 0.234  
& KRb  & 7.69 &  0.234 & 16.99 & -3.16 & 18,528 
 & 10.7 & 126
\\
${\cal Q}$ & 17.06  
& LiNa & 5.45 &  0.246 & 10.56 & -1.80 &  4,265 
 & 6.34 &  74.4
\\
${\cal O}$ & -23.71   
&   &     &        &        &            
&   &
\\
$C_{6,0}$ & 11679 
& RbCs & 8.37 &  0.554 & 14.19 & -5.39 & 26,599
 & 21.6 &  44.6
\\
$C_{6,1}$ & 3182  
&    &      &        &        &              
&    &
\\
$C_{6,2}$ & 10441   
& LiRb & 6.50 &  1.715 & 11.80 & -1.61 & 8,528        
 & 9.95 &  12.0
\\
$C_{6,3}$ &-2893  
& LiCs & 6.93 &  2.335 & 11.00 & -7.26 & 10,951      
& 12.7 &   8.5 
\\
$C_{6,4}$ & 158  
& NaK  & 6.61 &  1.199 & 12.91 &  3.83  & 9166         
 & 9.52 &  18.5
\\ \hline\hline
\end{tabular}
\caption{\label{tab1}
Left: fit parameters (up to $R^{-6}$). Right, {\it ab initio} values of the
equilibrium separation $R_e$, moments ${\cal D}$, ${\cal Q}$, and ${\cal O}$
(from the geometric center), $W_6$ for the collinear orientation, and the turning
points $R_{\rm sr}$ and $R_{\cal Q}$ for various molecules AB in $v=0$ of
X$^1\Sigma^+$.  All values are in atomic units.}
\end{table} 

\begin{figure}[b]
 \caption{ \label{fig4} $V_{\rm top}$ vs. $\theta_1$ and $\theta_2$. The main
 plot corresponds to a twist angle $\phi =0$, while the two smaller plots to
 $\phi = 40^{\rm o}$ (top) and $80^{\rm o}$ (bottom).  $V_{\rm top}$ is set to
 zero if there is no barrier. }
 \includegraphics[width=\columnwidth]{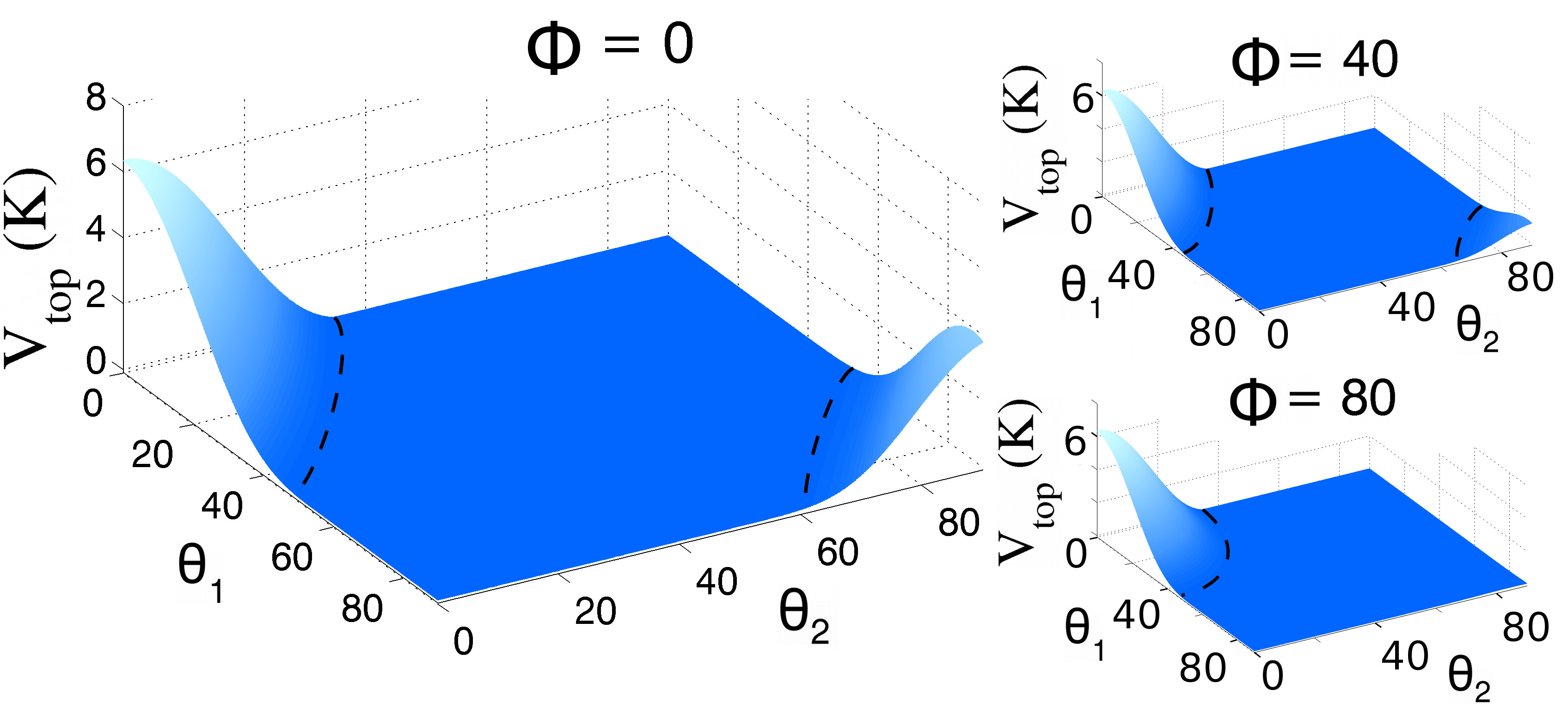}
\end{figure}

Using Eq. (\ref{eq:expansion-general}), we study
the geometries leading to a long-range barrier;
Fig.~\ref{fig4} depicts its height $V_{\rm top}$ 
as a function of $\theta_1$ and $\theta_2$ for a few twist angles $\phi$.
For $\phi =0$, a substantial barrier exists 
along the diagonal $\theta\equiv\theta_1 = \theta_2$, for small 
angles ($\theta \sim 20^{\rm o}$ or less), and for large angles 
($\theta \sim 70^{\rm o}$ or more). While the barrier remains present for
the small angle cone ($\sim 20^{\rm o}$) as $\phi$ increases, 
it quickly disappears for large $\theta$. Roughly speaking, there is a 
barrier for a cone of $\theta \sim 20^{\rm o}$ for any $\phi$, and for larger
molecular misalignment, the barrier vanishes. 
A significant barrier can thus be maintained by aligning 
the molecules within a small angular cone, allowing
ultracold KRb samples to remain stable and even be evaporatively cooled in 
various trap geometries (1-D when nearly collinear, and 1-D or 2-D
when nearly aligned).

\begin{figure}[t]
 \includegraphics[width=\columnwidth]{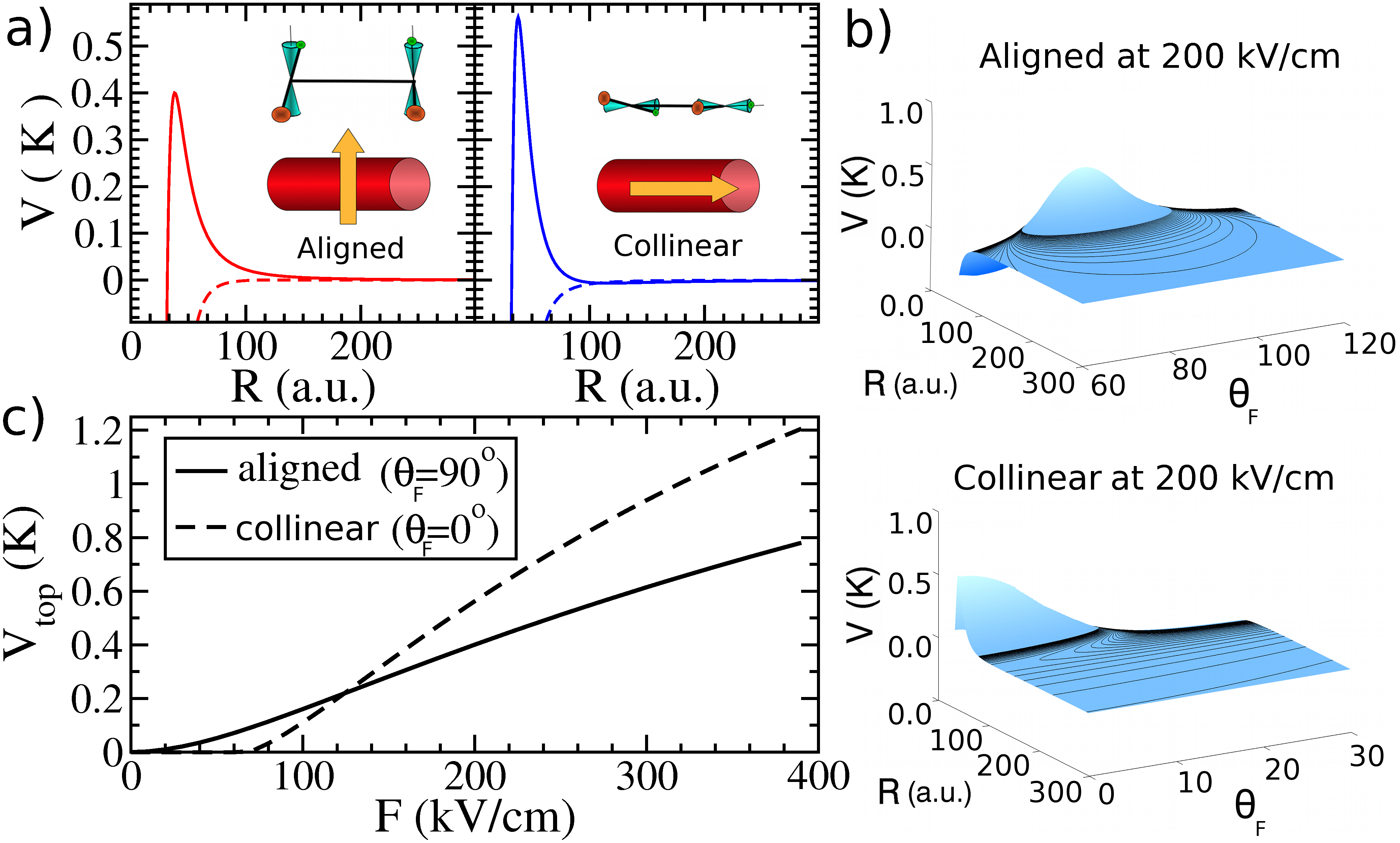}
 \caption{ \label{fig:elec-low-high} (a) KRb+KRb interaction (1-D) for weak (5
 kV/cm: dashed lines) and strong electric fields (200 kV/cm: solid lines),
 oriented perpendicular (left) and parallel (right) to the intermolecular axis.
 The red cylinder represents the 1-D trap, the arrow the orientation of the
 field, and the sketch above the precessing molecules.         (b)
 Intermolecular interaction with $F= 200$ kV/cm for the aligned (top) and
 collinear (bottom) geometries as a function of $\theta_F$ (angle between $\bf F$
 and $\bf R$).  (c) Height of the barrier for the aligned and collinear
 orientations as a function of $F$.}
\end{figure}

Polar molecules can be oriented by coupling rotational states along a polarizing
external electric field ${\bf F}$.  This can be achieved by using a DC electric field;
however the small dipole moment of KRb requires field strengths that are difficult to 
achieve in the laboratory.  An alternative is to add a separate polarizing laser
field \cite{hartelt2008} that directly couples the rotational states of the
molecule. Although this requires a much smaller DC field, non-adiabatic
effects are prominent \cite{nielsen2012}, and for the sake of simplicity 
we calculate the rotational state coupling through directly scaling the external field.
In the rigid-rotor approximation, we get a superposition of field-free symmetric top states 
\begin{equation}
|\tilde{J}\tilde{M}\Omega \rangle=\sum_{J,M}
a_{M,\tilde{M}}^{J,\tilde{J}}|JM\Omega \rangle
\end{equation}
labeled by their total angular momentum $J$ with projection $M$ along
${\bf F}$. After transforming the molecule-fixed frame
potential $V(R,\theta_1,\theta_2,\phi)$ to the laboratory-fixed frame $V_{\rm
Lab} ({\bf R},\hat{r}_1,\hat{r}_2)$ \cite{tscherbul2009}, the field averaged
potential is found by evaluating
\begin{equation}\label{labframe}
V({\bf R}) =
\langle\tilde{J}'\tilde{M}'\Omega'| V_{\rm Lab} ({\bf R},\hat{r}_1,\hat{r}_2)
|\tilde{J}\tilde{M}\Omega\rangle.
\end{equation}
In Fig.~\ref{fig:elec-low-high}, we illustrate the effect of ${\bf F}$ on a pair
of KRb molecules in 1-D, with $\theta_F$ defined as the angle between ${\bf F}$ and ${\bf R}$.  
For weak fields ($F\alt10$ kV/cm), the molecules remain largely in their $J=0$
rotational state, the field mixes only small amounts of higher $J$ states.
Classically, they precess ``wildly" on a wide cone about ${\bf F}$, and thus the
KRb+KRb interaction samples a large range of relative angles, averaging its
attractive and repulsive components, with the main contribution arising mainly 
from the isotropic attractive van der Waals $C_{6,0}$ term.  This
is depicted by the dashed lines in Fig.~\ref{fig:elec-low-high}(a) with a field of
5 kV/cm for the aligned (left) and collinear (right) orientations. In both cases,
the interaction becomes strongly attractive at short distance, with the aligned
geometry having a weak barrier ($\sim 1$ mK) and the collinear case
showing no sign of a barrier.  The solid lines 
show the effect of a larger electric field of 200 kV/cm; ${\bf F}$ strongly mixes
many ($\sim7$) more $J$'s and classically the molecules precess on a tighter cone,
sampling a more restrictive range of angles. The anisotropic interactions do not
average to zero, and strong barriers are present for both aligned and collinear
cases. Fig.~\ref{fig:elec-low-high}(b) shows the interaction for a range of $\theta_F$ 
near the aligned and collinear orientations for $F=200$ kV/cm. 
Since the molecules then behave almost like rigid rods, we recover results similar
to those in the molecular-frame.  The barrier survives
for a cone of angle $\theta_F$ of about 20$^{\rm o}$ for both orientations, and
the same conclusions about stability of 1-D and 2-D samples apply. For the aligned
orientation, the barrier appears rapidly even for low fields, while larger fields 
($F\agt70$ kV/cm) are necessary for the collinear case (see
Fig~\ref{fig:elec-low-high}(c)).  In both cases, the barrier grows rapidly to
hundreds of mK, a value much higher than the typical kinetic energy of the
trapped ultracold molecules ($ < 100$  $\mu$K).

\begin{figure}
 \includegraphics[width=\columnwidth]{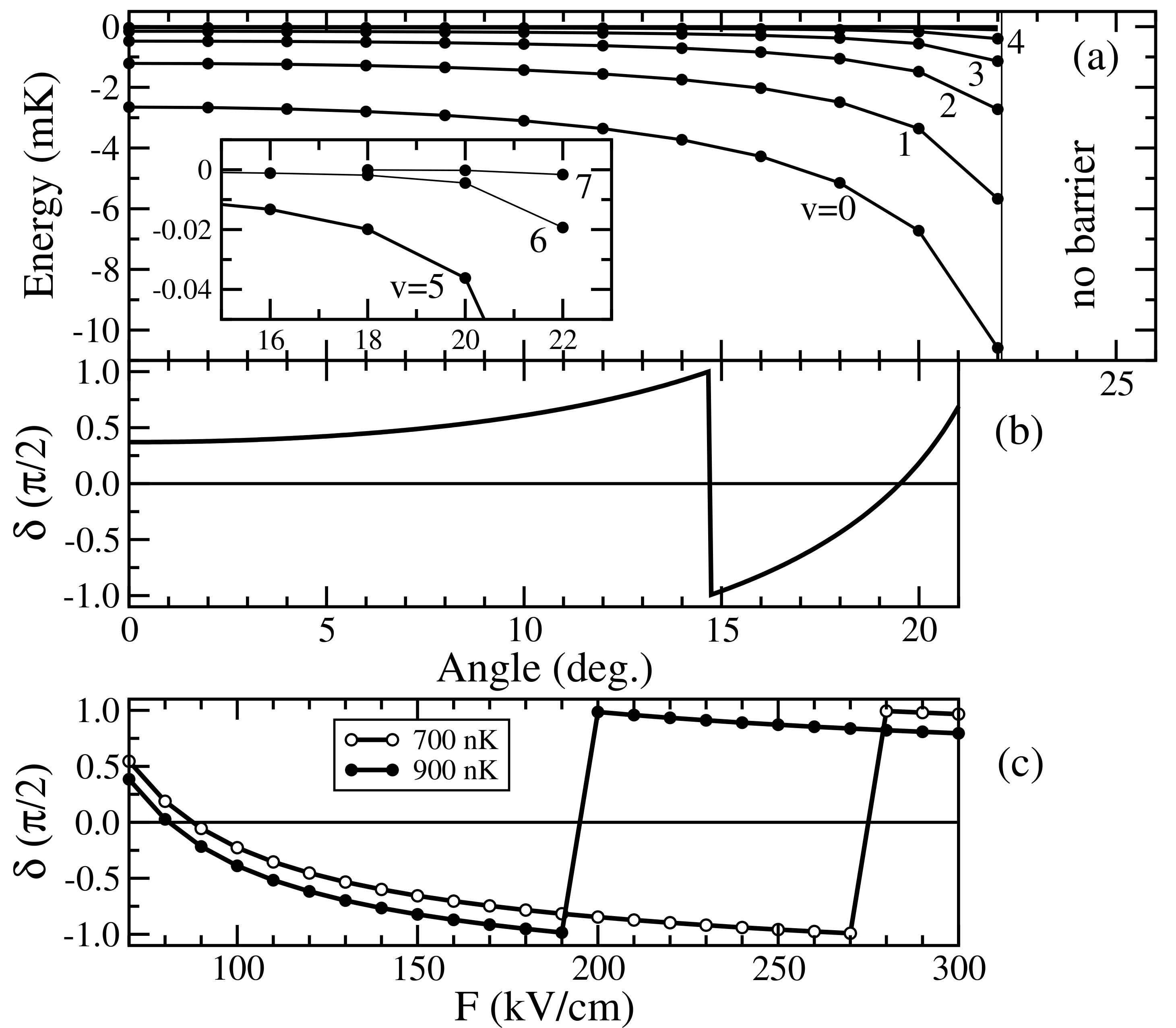}
 \caption{ \label{fig:delta-sigma-mod} (a) Long-range well energy levels vs.
 $\theta$; an additional level $v=7$ appears at 18$^{\rm o}$ (inset). (b)
 scattering phase shift $\delta$ vs. $\theta$ for $k$ corresponding to 900 nK
 for infinite $F$.  (c) $\delta$ for $\theta_F=0$ as a function of the field
 strength $F$ for collision energies corresponding to 700 nK and 900 nK.}
  \end{figure}

Fig.~\ref{fig:elec-low-high}(b) hints at the existence of a long-range well for the
collinear geometry. We analyze this well in the limit of infinite electric field, when
the molecules are fully parallel ($\theta\equiv\theta_1=\theta_2$ and  $\phi =0$) and
both molecular-frame and laboratory-frame (for 1-D trapped molecules) potentials 
coincide ($\theta=\theta_F$).
We find a long-range well that
can sustain several bound levels due to its large extension and the large mass
of the KRb molecules. For $\theta=0$, there are 7 levels, the deepest bound
by nearly 2.7 mK with inner and outer classical turning points at 110 a.u. and
205 a.u., respectively.  As $\theta$ increases, the barrier due to the $R^{-5}$
repulsion gets smaller and the well deepens, and the binding energies increase
accordingly until $\theta$ reaches $\theta_c \simeq 22^{\rm o}$, at which point
the barrier disappears and no more long-range bound levels exist 
(see Fig.~\ref{fig:delta-sigma-mod}(a)). We note that
for a small deviation from $\theta =0$, the binding energies are not
significantly affected, and an additional 
level $v=7$ appears for $18^{\rm o}<\theta < \theta_c$ (inset in Fig.~\ref{fig:delta-sigma-mod}(a)).

The variation of bound levels with $\theta$ affects the scattering between
molecules and their effective interaction.  Assuming $\theta$ (or $\theta_F$) as a
fixed external parameter, we estimate the $s$-wave scattering phase shift
$\delta$ between two KRb, which depends on the interaction $V$ and the
wave number $k$; $\delta<0$ ($>0$) corresponds to an effective
repulsive (attractive) interaction.  Here, we choose $k$ assuming 
$\hbar^2 k^2 \sim m k_B T$ ($m$: mass of KRb; $k_B$: Boltzmann constant) 
for $T\simeq 700$ nK \cite{miranda2011} and
illustrate the effect in Fig.~\ref{fig:delta-sigma-mod}(b) for the infinite field case
({\it i.e.} $\theta_F=\theta$).  For angles smaller
than $\sim 14.7^{\rm o}$, the interaction is attractive (with $\delta > 0$),
while it becomes repulsive ($\delta < 0$) for larger angles.  
In an ideal 1-D trap, the repulsive barrier at $R\sim 100$ a.u. would stabilize
the sample for an attractive effective interaction by preventing the molecules from 
reaching short distances where inelastic processes ({\it e.g.}, chemical reactions) 
could take place. Larger angles, but still within the stability cone, would also give
stable samples since the effective interaction is repulsive.
By varying the orientation of the electric field with respect to the trap axis, 
the behavior of the sample could be controlled; an effective 
attractive interaction would lead to a dense self-trapped system, 
{\it i.e.} a liquid-like sample, while an effective repulsive interaction would
give a dilute sample behaving like a gas.  Such control could probe a 
quantum phase transition between a Luttinger liquid and an ultracold 
gas \cite{recati2003}.  One could also create a
chain of KRb molecules weakly bound together ({\it e.g.} by 
using photoassociation); these would be akin to
ultracold polymer-like chains stabilized by an external electric field and a 1-D
trap. We note that the effective interaction can also be controlled by
varying the magnitude of ${\bf F}$. In Fig.~\ref{fig:delta-sigma-mod}(c),
we show $\delta$ for $\theta_F=0$ as a function of $F$ for two collision energies 
corresponding to 700 nK and 900 nK, and find that its sign can be changed by varying $F$.

Obtaining 1-D traps is challenging; assuming a harmonic trap in the perpendicular 
direction characterized by the frequency $\omega$, the size of the ground state 
$a\sim \sqrt{\hbar/m\omega}$ is of the order of a few 1000 a.u. for optical traps.
Molecules at densities of $10^{12}$ cm$^{-3}$ loaded in such traps would be 
separated by roughly $d\sim 1$ $\mu$m, and for repulsive effective interaction, the
angle $\tan^{-1}a/d \alt 10^{\rm o}$ between their axes would remain within the 
cone of stability. For an attractive effective interaction, the relevant angle is 
$\tan^{-1}a/R_{\cal Q}$, where $R_{\cal Q}$ is the point where the barrier
begins for two approaching molecules (see below), which requires $a\sim 0.4
R_{\cal Q}= 50$ a.u. for KRb.
Here, the sample would not be 1-D, with inelastic processes possibly taking place.
Non-reactive species, such as RbCs (see below), could be considered to prevent
inelastic processes, or much tighter magnetic traps could be employed; in which case
molecules in a triplet electronic state with a magnetic moment $\mu$ would be 
required. For KRb in its a$^3\Sigma^+(v=0)$ state, $a\sim 60$ a.u. can be
achieved \cite{schmiedmayer2002}, and with $R_{\cal Q}\sim 150$ a.u.
\footnote{${\cal D}=0.017$ a.u. and ${\cal Q}=1.47$ a.u. were computed at the same level
of theory mentioned in the text, leading to $R_{\cal Q} \sim 150$ a.u. for KRb.},
$\tan^{-1}a/R_{\cal Q}$ would remain within the stability cone.

The features discussed here for KRb can be generalized to other polar molecules.
The PES at long-range is well described by (\ref{eq:expansion-general}), and the
existence of a barrier for perfectly collinear molecules depends mostly on the
first two terms (see Eq.(\ref{eq:V-approx})).  By setting $V=0$ and neglecting
${\cal DO}$, we find $R_{\cal Q}\simeq \sqrt{3{\cal Q}^2/{\cal D}^2}$, the point where
the $R^{-5}$ repulsion takes over the $R^{-3}$ attraction.  We can also define
a point $R_{\rm sr}$ where the shorter range $R^{-6}$ attraction takes over the
$R^{-5}$ repulsion, by neglecting the other contributions and setting
$V\sim -W_5/R^5-W_6/R^6 = 0$, which gives $R_{\rm sr}=-W_6/W_5$.
If $R_{\cal Q}$ is outside the region where bonds are strongly perturbed 
($\sim 20$ a.u.) or higher $W_n$ terms contribute significantly (roughly $R_{\rm sr}$), 
then the barrier can exist. Table~\ref{tab1} gives $R_{\cal Q}$ and $R_{\rm sr}$ 
for various polar molecules. Because ${\cal Q}$
has roughly the same amplitude for most of them, ${\cal D}$ dictates the
behavior of the systems. Molecules with small ${\cal D}$ ({\it e.g.}, LiNa and
KRb) have a sizable $R_{\cal Q}$, and thus the existence of a barrier is very likely,
unlike those with a large ${\cal D}$ ({\it e.g.}, LiRb, LiCs and NaK). We also
include RbCs, for which the existence of a barrier is uncertain. This is
interesting since RbCs is known to not be reactive. However, a full
investigation is required to know if a barrier exists.

In conclusion, we found that the interaction between polar molecules exhibits a
strong barrier when they are oriented about two specific geometries: aligned and
collinear. We also showed that the collinear setting gives meta-stable samples
of ultracold molecules in a tight 1-D trap. The long-range $R^{-3}$ dipolar
attractive and $R^{-5}$ quadrupolar repulsive contributions in the collinear
geometry lead to long-range wells between polar molecules sustaining several
bound levels. Varying the orientation of the molecules using an external
electric field allows for non-trivial effects, such as changing the effective
interaction from repulsive to attractive, and possibly the phase of the sample
from gas to liquid. Finally, we also predict the existence of the collinear
barrier for various bi-alkali polar molecules based on the relative strength of
the dipole and quadrupole moments.  The combination of available techniques to
produce ultracold molecules \cite{miranda2011,deiglmayr2011}  and the ability to
precisely control their spatial orientation \cite{nielsen2012,holmegaard2009}
provide the tools to investigate such systems.

The authors wish to thank H. Harvey Michels and J\"org Schmiedmayer for useful discussions.
This work was supported in part by the Department of Energy, Office of 
Basic Energy Sciences and the AFOSR MURI grant on ultracold polar molecules.

%\bibliography{library}

%Merlin.mbs v4.21 2009-07-09.
%

\end{document}